# Magnetically dead layer in interacting ultrafine NiFe$_2$O$_4$ nanoparticles

Yu. V. Knyazev[1*], D. A. Balaev[1], S. V. Stolyar[1,2], A. O. Shokhrina[2], D. A. Velikanov[1], A. I. Pankrats[1], A. M. Vorotynov[1], A. A. Krasikov[1], S. A. Skorobogatov[1], M. N. Volochaev[1], O. A. Bayukov[1], and R. S. Iskhakov[1]

[1]*Kirensky Institute of Physics, Federal Research Center KSC SB RAS, Krasnoyarsk, 660036 Russia*
[2]*Federal Research Center KSC SB RAS, Krasnoyarsk, 660036 Russia*
[*]yuk@iph.krasn.ru

**Abstract**—The interplay of the magnetically dead layer and structural defects in interacting ultrafine NiFe$_2$O$_4$ nanoparticles (<$d$> = 4 nm) have been investigated using transmission electron microscopy, X-ray diffraction, $^{57}$Fe Mössbauer spectroscopy, and dc magnetization and ac susceptibility measurements. According to the magnetic measurement data, there are three magnetic subsystems in NiFe$_2$O$_4$ nanoparticles. The first subsystem with the lowest blocking (spin freezing) temperature ($T_S$ = 8 K) involves atomic magnetic moments of magnetically disordered particles with a size of $d$ < 4 nm. The other two subsystems are formed by magnetic moments of the cores of nanoparticles more than 4 nm in size and by correlated surface spins in nanoparticle clusters. The magnetic moments of the ferrimagnetically ordered cores are blocked at a higher temperature (~40 K). It has been shown that the most significant contribution to the energy dissipation is made upon blocking of the correlated nanoparticle surface spins from the magnetically dead layer on the nanoparticle surface. The magnetic measurements have shown that the thickness of this layer is $d_{md}$ ≈ 1 nm for a particle with a diameter of <$d$> ≈ 4 nm. At the same time, the $^{57}$Fe Mössbauer spectroscopy study has revealed a structural disorder penetrating to a depth of up to $d_{cd}$ ≈ 0.6 nm in a particle with a diameter of <$d$> = 4 nm. This evidences for a faster violation of the magnetic order than in the case of the crystal order upon moving away from the center of a particle to its periphery.

## 1. Introduction

The increasing interest in magnetic nanoparticles is due to their high potential for use in various fields, including biomedicine, catalysis, ecology, and others. However, each field of application requires certain physical characteristics of nanoparticle systems used. In view of this, there is a need for finding ways of obtaining new nanomaterials and studying available ones. Obviously, the magnetic properties of nanoparticles and bulk materials can be essentially different, so it is often difficult to forecast the magnetic characteristics of a compound consisting of nanoparticles of a specific size. Therefore, it becomes important to build a model of the magnetic state of nanoparticles using experimental results. Such a model can help identify a key factor determining the observed modification of the magnetic properties of magnetic nanoparticles. In fact, it should be clearly understood how surface and size effects will change the magnetic characteristics of nanoparticles.

Spinel ferrites are distinguished among various oxide nanomaterials by their remarkable magnetic properties and relative ease of synthesis and modification [1]. In these materials, the saturation magnetization and bulk magnetic anisotropy constant can be changed by partial replacement of 3$d$ elements, which makes it possible to create particles with the magnetic characteristics required for a specific application (see, for example, reviews [1,2]). Nickel ferrite discussed in this work has a fairly high Néel temperature and the saturation magnetization of bulk NiFe$_2$O$_4$ is $M_S$ ~ 50 emu/g. This makes NiFe$_2$O$_4$ nanoparticles promising candidates for use in various fields [1–5].

The saturation magnetization of nickel ferrite nanoparticles decreases with their size [6–13]. This is a common property of many iron oxide nanoparticles [14–21]. The main factor reducing the $M_S$ value is structural defects localized on the nanoparticle surface, the role of which increases with decreasing particle size. These defects destruct the ferrimagnetic order, thereby making a negative contribution to the total magnetization. Here, it should be noted that a ferrimagnetic nanoparticle involves at least two magnetic subsystems: a ferrimagnetically ordered particle core and a surface spin subsystem [9–11,18–20,22,23]. However, despite all the rationality of this model of the particle magnetic state, the problems of estimating sizes of the above-mentioned magnetic subsystems and studying their behavior and magnetic response in different temperature ranges remain unsolved.

Certainly, the increasing role of surface states is not the only factor affecting the magnetic properties of nanomaterials. A decrease in the size of a granular magnet naturally leads to an increase in the number of defects in a grain and, consequently, to a decrease in the saturation magnetization of a sample [24]. In ultrafine (2–6 nm) particles of antiferromagnetic materials, structural defects can cause partial violation of the antiferromagnetic order and the occurrence of an uncompensated magnetic moment of particles. In antiferromagnetic particles, the increasing structural imperfection leads to an increase in the value of the uncompensated magnetic moment of particles, which becomes comparable with the magnetic moment of ferrimagnetic particles of similar sizes [25–30].

The occurrence of defects in the crystal structure enhances the manifestation of surface effects. In recent works [31,32], a detailed examination of the so-called magnetically dead layer in nanoparticles was reported. The authors analyzed the change in the saturation magnetization of core–shell nanoparticles by considering the disordered surface layer of the magnetic moments and the role of surface defects. In [33], it was shown using the small-angle neutron scattering examination of single-domain magnetite nanoparticles that the latter are divided into two magnetic subsystems, being still structurally homogeneous. The reported thickness of the magnetically dead layer was 1–1.5 nm at an average nanoparticle diameter of 9.0 nm. This feature becomes increasingly pronounced with decreasing nanoparticle size, which has recently been observed on ferrihydrite nanoparticles [25,26].

However, the separation of possible structural and magnetic disorder is still an urgent task in studying such objects. A detailed study of the magnetic and crystallographic disorder in such fine objects is extremely important for both fundamental research and application. First of all, this is required by the magnetic hyperthermia field intensively developed in recent years [34,35]. To correctly determine therapy parameters, the SAR of such drugs must be accurately estimated [36] in order to prevent unwanted damage to body tissues. Therefore, the quantitative estimation of the dependence of the surface imperfection of nanoparticles on their size is a crucial issue. This fact has already been noted in the literature [37], but so far serious studies in this direction have been lacking.

To meet this challenge, a deep and comprehensive analysis of the magnetic properties is needed, which should be consistent with the microstructure examination and make it possible to unambiguously divide contributions of the magnetic subsystems formed in magnetic nanoparticles. In this work, the magnetic properties of a powder system of nickel ferrite nanoparticles with an average size of ~4 nm and a narrow size distribution (the maximum size is no more than 6 nm) were studied. In such fine particles, the surface and size effects influence the magnetic state the most. One cannot ignore the presence of defects in the crystal structure of fine particles, which has a drastic impact on the magnetic properties. The aim of this study was to build a model of the magnetic state of the obtained $NiFe_2O_4$ nanoparticle system using the results of static magnetometry (the magnetization curves and temperature dependences of the magnetization), ac magnetic susceptibility measurements, and $^{57}Fe$ Mössbauer spectroscopy in combination with the transmission electron microscopy (TEM) and X-ray diffraction (XRD) data.

## 2. Experimental

Nickel ferrite nanoparticles were synthesized using the following chemical co-precipitation process: 0.2 mol $NiSO_4 \cdot 7 H_2O$ and 0.4 mol $FeCl_3 \cdot 6 H_2O$ were dissolved in 100 ml $dH_2O$, added dropwise with $NH_4OH$ to pH 11, and heated to 80°C for 30 min under continuous stirring. The resulting precipitate was washed and collected for investigations [34].

X-ray diffraction patterns were obtained on a HAOYUAN DX-2700BH diffractometer ($\lambda$ = 0.154 nm). The TEM and microdiffraction examination was carried out on a Hitachi HT7700 transmission electron microscope (the accelerating voltage is 100 kV) with an energy-dispersive X-ray spectroscopy (EDS) microanalysis system. Specimens were prepared by shaking the nanoparticle powder in alcohol in an ultrasonic bath and depositing the obtained suspension onto support meshes with a perforated carbon coating. The time of accumulation for the EDS analysis was determined by the spectrum assembly quality, which allows for the quantitative processing and was no shorter than 10 min. The phase composition of the sample in a local field was determined by the selected-area electron diffraction (SAED) technique.

The Mössbauer spectra of the sample were obtained on an MS-1104Em spectrometer (the Research Institute of Physics, Southern Federal University) in the transmission geometry with a $Co^{57}(Rh)$ radioactive source in the temperature range of 4–300 K using a CFSG-311-MESS cryostat with a sample in the exchange gas based on a closed-cycle Gifford-McMahon cryocooler (Cryotrade

Engineering, LLC). The spectra were processed by varying the entire set of hyperfine parameters using the least squares method in the linear approximation. The spectra were fitted by Lorentzian lines, taking into account the broadening caused by the magnetic and crystalline inhomogeneity of the sample.

The $M(H)$ curves and $M(T)$ dependences were measured using a vibrating sample magnetometer [38]. The $M(T)$ dependences in a low field ($H$ = 2 Oe) were measured using a SQUID magnetometer [39]. The temperature dependences of the magnetization were obtained in the zero field cooling (ZFC) and field cooling (FC) modes. The temperature dependences of the real ($\chi'$) and imaginary ($\chi''$) parts of the ac magnetic susceptibility were measured using a Quantum Design Physical Property Measurement System (PPMS-9) at frequencies of 10, 100, 1000, and 10000 Hz with an ac field amplitude of 2 Oe.

## 3. Results
### *3.1 TEM and XRD Studies*

Figure 1a shows a typical TEM image. The particle size determined from several microphotographs ranges within 2–6 nm, so the average size is $<d> \approx 4$ nm. A size distribution histogram is presented in the inset to Fig. 1a. Typical SAED patterns contain two distinct diffraction rings. This is consistent with the XRD pattern (Fig.1b), in which only two corresponding reflections can be clearly seen. In addition, the Bragg peaks are strongly broadened, which is indicative of a low degree of crystallinity characteristic of nanoparticles. The observed diffraction peaks correspond to the highest-intensity peaks of the $NiFe_2O_4$ structure with the (113) and (044) Miller indices.

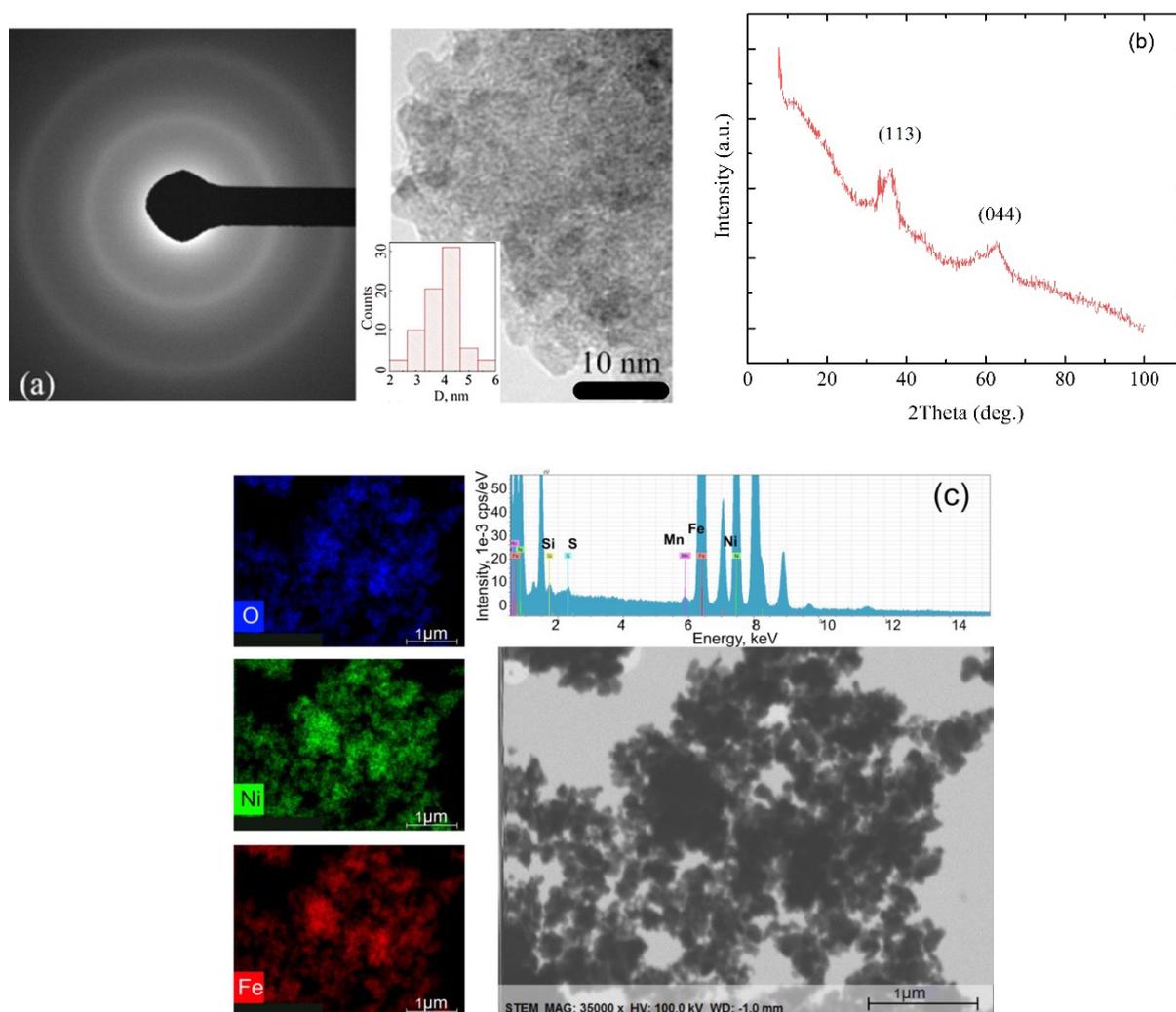

Fig. 1. (a) SAED pattern and corresponding TEM image of the sample and (b) XRD data. Miller indices for $NiFe_2O_4$ are given next to the X-ray peaks [40]. (c) EDS mapping and X-ray absorption spectrum of $NiFe_2O_4$ nanoparticles.

The coherent scattering region size $d_{CS}$ can be calculated from the XRD data using the Debye–Scherrer formula [41]

$$d_{CS} = \frac{K_{SH} \cdot \lambda}{\beta \cos \Theta} \qquad (1)$$

Here, $K_{SH}$ is the particle shape coefficient, $\lambda$ is the X-ray wavelength (CuKα radiation with a wavelength of 1.541Å), β is the full width at half maximum of the reflection obtained from the data shown in Fig. 1, and Θ is the Bragg angle (rad) of the corresponding peak. The coefficient $K_{SH}$ for spherical particles is usually taken to be 0.9. The estimation yielded an average size of $d_{CS} \sim 3.5$ nm, which is much smaller than the <d> value obtained by analyzing the nanoparticle size using the micrographs.

The results of the elemental mapping performed on an area of 5 × 3.5 μm (Fig. 3c) show the high intensity of iron and nickel atoms. According to the EDS data, the Fe, Ni, and O concentrations are 30±4, 14±2, and 56±2 at %, respectively. Using these atomic ratios, one can determine the chemical formula of the investigated nanoparticle ensemble, which is $NiFe_{2.1}O_4$ in the oxygen stoichiometry approximation. This formula is similar to the chemical formula of spinel; the slight deviation can be attributed to an error of the experimental determination of the atomic fraction.

*3.2 Mössbauer Effect*

Figure 2 presents the Mössbauer spectra obtained at temperatures of 300 and 4 K. At room temperature, the spectrum is a quadrupole doublet, which can reflect both the superparamagnetic (SPM) state of the magnetic moments of $NiFe_2O_4$ nanoparticles and the paramagnetic (PM) state of individual spins [42,43]. At 4 K, the spectrum demonstrates the Zeeman splitting of the Mössbauer lines, which corresponds to the state with the blocked magnetic moments of nanoparticles.

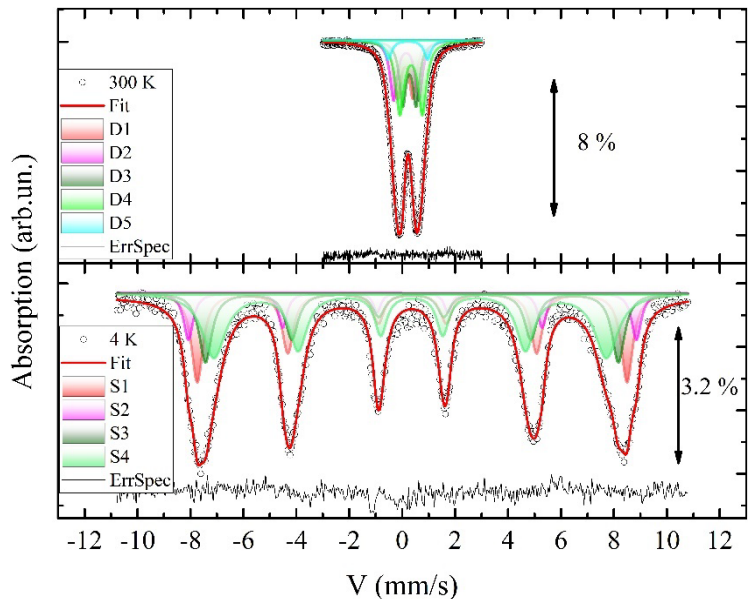

Fig. 2. Mössbauer spectra of $NiFe_2O_4$ nanoparticles at temperatures of 300 K (on the top) and 4 K (in the bottom). Dots show the experimental spectra and the solid line, the result of processing. Shaded areas correspond to the spectral components. The difference spectrum of the error is shown under the Mössbauer spectrum. The value of the effect is indicated to the right of the spectrum.

The results of processing of the spectra are given in Table 1. The chemical shift δ of partial doublets points out the iron charge state 3+. The δ value is sensitive to the local environment of iron atoms; therefore, the $^{57}$Fe environment type for the spinel structure can be easily determined [44]. According to the results obtained by us, iron is in both the octahedral (D3, D4, D5) and tetrahedral (D1, D2) environment. Since the hyperfine parameter Δ indicates the degree of local distortions in the SPM/PM state, it can be noted that iron cations, which are mathematically described by the quadrupole doublet D5, should be classified as the most distorted states of iron that can form on the particle surface. In addition, it can be noted that the pair of doublets D1 and D3 has much smaller quadrupole splitting values than the pair of doublets D2 and D4. Therefore, it is obvious that the doublets D1 and D3 belong to iron cations with the more symmetric environment, while the doublets D2 and D4, on the contrary, to the state with stronger local distortions. Under that logic, iron in the nanoparticles under study has the environment of two types with different degrees of crystallization.

Table 1. Mössbauer parameters of the $NiFe_2O_4$ sample at temperatures of 4 and 300 K. $\delta$ is the chemical shift relative to α-Fe, $H_{hf}$ is the hyperfine field on iron nuclei, $\Delta$ is the quadrupole splitting, $W$ is the Mössbauer line FWHM, and $A$ is the relative site occupancy. Positions with a high level of the local distortions are italicized.

|    | $\delta$, mm/s ±0.005 | $H_{hf}$, kOe, ±3 | $\Delta$, mm/s ±0.01 | $W$, mm/s ±0.01 | $A$, a. u. ±0.03 | Origin |
|----|---|---|---|---|---|---|
|    |   |   | 300 K |   |   |   |
| D1 | 0.221 | -- | 0.58 | 0.33 | 0.19 | $Fe^{3+}$-tetra |
| *D2* | *0.258* | -- | *0.95* | *0.34* | *0.22* | *$Fe^{3+}$-tetra* |
| D3 | 0.380 | -- | 0.50 | 0.31 | 0.21 | $Fe^{3+}$-octa |
| *D4* | *0.445* | -- | *0.85* | *0.39* | *0.31* | *$Fe^{3+}$-octa* |
| *D5* | *0.324* | -- | *1.48* | *0.32* | *0.07* | *$Fe^{3+}$-octa* |
|    |   |   | 4 K |   |   |   |
| *S1* | *0.480* | *505* | *-0.01* | *0.53* | *0.26* | *$Fe^{3+}$-octa* |
| S2 | 0.489 | 526 | -0.00 | 0.34 | 0.12 | $Fe^{3+}$-octa |
| S3 | 0.458 | 485 | 0.05 | 0.57 | 0.23 | $Fe^{3+}$-tetra |
| *S4* | *0.435* | *461* | *-0.12* | *0.46* | *0.39* | *$Fe^{3+}$-tetra* |

Thus, the quadrupole doublets D1 and D3 with smaller splitting values reflect the well-crystallized areas of ferrite nanoparticles, while the doublets D2, D4, and D5 may correspond to the regions with the worse crystallization (with more defects) and to the surface states (D5). The local structural and magnetic order of the nanoparticle areas can be estimated from the deviation of the hyperfine field value on iron nuclei at 4.2 K (shown in italics in Table 1). This parameter is highly sensitive to a number of magnetic bonds and its drop is indicative of the presence of defects in the local environment. In addition, structural defects lead to local structural distortions; this is reflected in the Mössbauer spectra as a growth of the quadrupole splitting of quadrupole doublets at 300 K. Therefore, using the spectrum component areas obtained, one can calculate the atomic ratio of iron in the "good" and "bad" nanoparticle areas in terms of symmetry of the local environment. This will be expressed through the ratio between the areas of the above-mentioned doublets: 40% and 60%, respectively. Using the analogous data obtained at 300 K, we can write the crystal chemical formula for the ferrite nanoparticle areas of both types, taking into account the existence of the poorly and well crystallized regions:

$(Fe_{0.95}Ni_{0.05})[Fe_{1.05}Ni_{0.95}]O_4$ for the "good" areas,
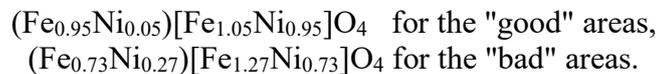
$(Fe_{0.73}Ni_{0.27})[Fe_{1.27}Ni_{0.73}]O_4$ for the "bad" areas.

These formulas allow us to estimate the magnetic moment per formula unit of the investigated $NiFe_2O_4$ nanoparticle sample, taking into account the weighting coefficients of the nanoparticle areas. In the "good" areas, the magnetic moment is expected to be ~2.3$\mu_B$, while in the «bad» ones, ~3.6$\mu_B$. Thus, the average moment on the sample will be ~3.1$\mu_B$. However, as shown below, the observed magnetic moment is much smaller than that estimated from the Mössbauer spectroscopy data, which may be due to the strong degree of magnetic disorder in the system under study.

Now, let us turn to the low-temperature spectrum. Mathematically, it is modeled by four sextets (Table 1), which also indicate a partitioning into two possible nanoparticle types or nanoparticle areas. It can be noted that the widths of the Mössbauer lines of sextets do not broaden, which evidences for the homogeneity of the $^{57}Fe$ magnetic environment in nanoparticles. In this case, however, one should pay attention to the value of the hyperfine magnetic field on $^{57}Fe$ nuclei. This hyperfine parameter is highly sensitive to the magnetic environment of a cation. Since an increase in the defectiveness of the structure leads to a decrease in a number of magnetic couplings, the hyperfine magnetic field for iron cations with the defective environment will also decrease. Thus, the ratio between the numbers of iron atoms in the more and less ordered local environments can be determined. These are the sextets S2, S3 and S1, S4, respectively. The ratio between these pairs is 35% : 65%, which, with allowance for the processing error, almost coincides with the ratio obtained at 300 K.

*3.3 Magnetic Measurement Data*

The transition from the blocked to SPM state upon temperature variation, which is related to the magnetic moment $\mu_P$ of a single-domain particle, is manifested, as a rule, in the $M(T)_{ZFC}$ maximum at the SPM blocking temperature and the effect of thermomagnetic history below this temperature. The temperature dependences of the magnetization below 100 K in different fields and at different thermomagnetic prehistories (ZFC and FC) are presented in Fig. 3a. Under the ZFC conditions, the $M(T)$ dependence taken in fields of $H$ = 100–1000 Oe has two pronounced maxima at temperatures of $T_{max} \approx 40$ K and $T_S \approx 8$ K. The irreversible behavior of the magnetization observed in the low-temperature region (the difference between the $M(T)_{ZFC}$ and $M(T)_{FC}$ dependences) begins in the vicinity of the high-temperature maximum.

The presence of two $M(T)_{ZFC}$ maxima, together with the effect of the thermomagnetic prehistory (specifically, the relative positioning of the $M(T)_{ZFC}$ and $M(T)_{FC}$ curves) indicates the occurrence of complex SPM blocking (or freezing) processes at least in two magnetic subsystems. If we consider a nanoparticle within the core–shell paradigm (see Introduction), the high-temperature maximum can be attributed, in the first approximation, to blocking of the magnetic moments of particles and the low-temperature maximum, to freezing of the surface spins or, as shown in Section 4, to freezing of spins in the finest particles. A separate process of freezing of spins at $T_S$ is additionally indicated by a significant impact of the thermomagnetic prehistory, in particular, when field cooling is carried out in the range intermediate between the temperatures of two $M(T)_{ZFC}$ maxima. Figure 3a shows the $M(T)_{FC}$ dependence for which field cooling was carried out starting with 15 K. In this case, the $M(T)_{FC}$ curve exhibits a temperature behavior similar to that of the $M(T)_{FC}$ curve built after field cooling from a temperature above $T_{max}$, which suggests that the characteristic bell in the $M(T)_{FC}$ plot around the temperature $T_S$ is unrelated to the SPM blocking (freezing) processes, which occur starting with $T_{max}$. Hence, the low-temperature $M(T)_{ZFC}$ and $M(T)_{FC}$ maxima can be interpreted as the manifestation of freezing of spins on the particle surface or spins of the finest particles.

As the external field increases, the high-temperature maximum shifts towards lower temperatures and we state a fairly strong effect of the magnetic field on the $T_{max}$ value. In addition, it should be noted that, at $H \approx 0.5$–1.5 kOe, the $M(T)_{ZFC}$ dependence is plateau-shaped in the vicinity of $T_{max}$. The pronounced $M(T)_{ZFC}$ plateau in certain field and temperature ranges can be attributed to the fact that the energy of interaction between the magnetic moments of neighboring particles prevails over the thermal and Zeeman ($\mu_P \cdot H$) energy and, therefore, the total magnetization barely changes with temperature [45–47]. The two aforementioned features of the $M(T)$ curves were observed for the systems of nanoparticles under fairly strong magnetic interactions [45–48]. In other words, in the powder $NiFe_2O_4$ nanoparticle system studied here, the interparticle magnetic interactions also significantly contribute to the resulting magnetic properties. Analysis of the ac magnetic susceptibility (see Subsection 3.4) confirms this statement.

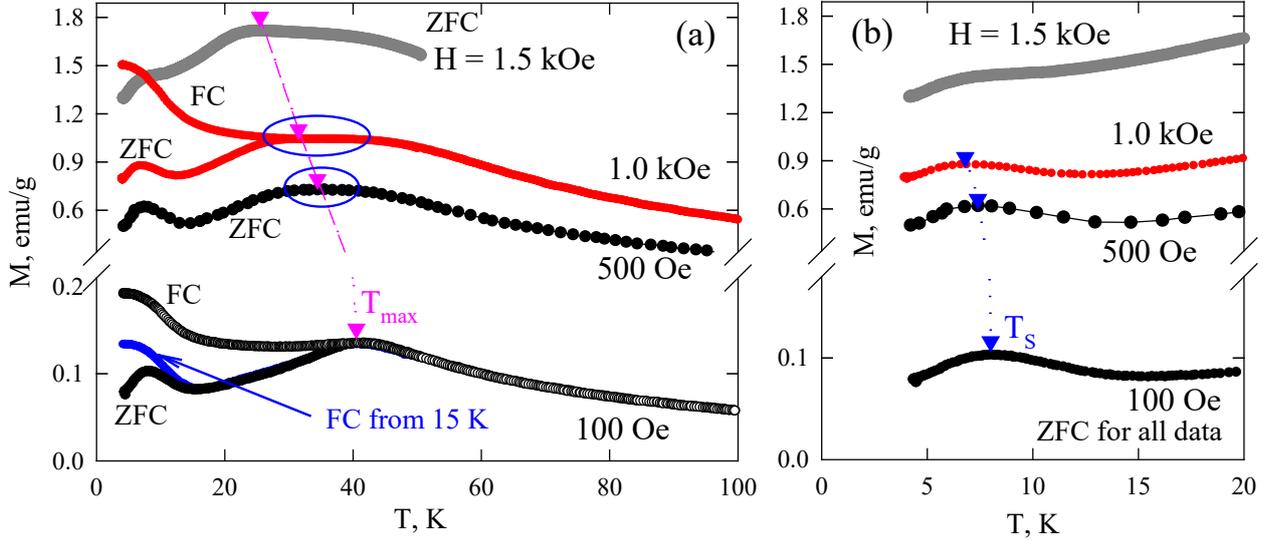

Fig. 3. ZFC and FC temperature dependences of the magnetization in different external fields. The indicated temperatures $T_{max}$ correspond to the $M(T)_{ZFC}$ maximum (blocking of the magnetic moments of particles) and the temperatures $T_S$, to the low-temperature $M(T)_{ZFC}$ maxima (freezing of the surface spin subsystem).

In contrast to the impact of the field strength on the $T_{max}$ position, the shift of the low-temperature maximum $T_S$ with increasing external field is weak (Fig. 3b); however, in a field of $H = 1.5$ kOe, the $M(T)_{ZFC}$ dependence contains no low-temperature maximum. This is not caused by a decrease in $T_S$ below 4.2 K, which can be seen from the evolution of the shape of the $M(T)_{ZFC}$ curve in the low-temperature region (Fig. 3b). Near 12–15 K, the $M(T)_{ZFC}$ dependences in fields of 100, 500, and 1 kOe have local minima, while in fields of 1.5 kOe and stronger, the minimum is not observed. Possibly, both the high- and low-temperature maxima can be described by two almost independent bell-shaped functions, which evolve with increasing external field. At the minimum point in weak fields, the contribution of the high-temperature bell to the total magnetization is still minor. As the external field increases, the high-temperature maximum shifts to the low-temperature region and the left-hand side of the bell (the dependence decreasing with temperature) passes to the region of the low-temperature maximum. Therefore, instead of the maximum around 4–10 K in fields of ~1.5 kOe and stronger, one can see an arc-shaped feature against the background of the $M(T)_{ZFC}$ dependence decreasing with temperature.

At $T = 4.2$ K, the $M(H)$ dependence is hysteretic; the coercivity (in the maximum applied field of 60 kOe) is ~2.7 kOe. To further analyze the magnetic characteristics, it is necessary to determine the contribution of the particle magnetic moments to the total magnetization. Figure 4a presents the $M(H)$ dependences for the investigated sample in the temperature range of 100–300 K. This temperature range intentionally exceeds the magnetic blocking temperatures (Fig. 3); therefore, the $M(H)$ dependences in Fig. 4a are reversible. The SPM behavior of the magnetization is usually modeled by the Langevin function $L(\mu_P, H) = \coth(\mu_P \cdot H/kT) - 1/(\mu_P \cdot H/kT)$, where $k$ is the Boltzmann constant. In our case, $\mu_P$ is obviously determined by the ferrimagnetic ordering in the NiFe$_2$O$_4$ compound. However, as can be seen from the general form of the $M(H)$ dependences, there is a significant contribution of the field-linear magnetization response. Therefore, the magnetization curves should be described using superposition of the Langevin functions and the field-linear function.

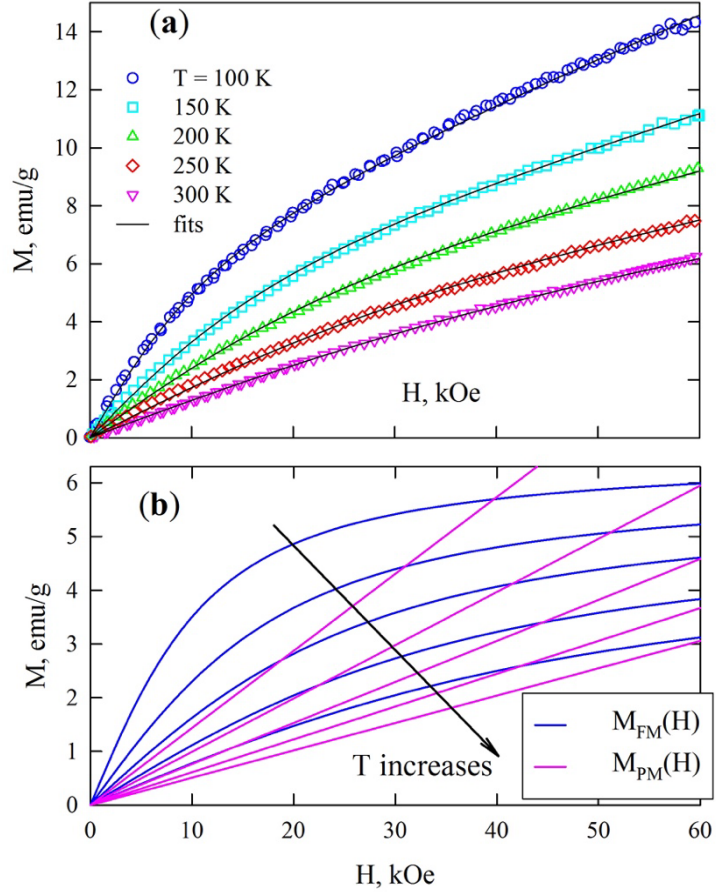

Fig. 4. (a) Experimental *M(H)* dependences (symbols) at the temperatures indicated in the legend and results of the best fitting using Eq. (2). (b) Partial components (the first ($M_{FM}(H)$) and second ($M_{PM}(H)$) terms of Eq. (2)) of the fitting curves.

To describe the experimental *M(H)* dependences, we use the standard equation

$$M(H) = N_P \int_0^\infty L(\mu_P, H) f(\mu_P) \mu_P \, d\mu_P + \chi_{PM} H, \qquad (2)$$

where $N_P$ is the number of particles per unit sample mass, $f(\mu_P)$ is the distribution over the particle magnetic moments, and $\chi_{PM}$ is the field-independent factor. As $f(\mu_P)$, the log-normal distribution $f(\mu_P) = (\mu_P \cdot s \cdot (2\pi)^{1/2})^{-1} \exp\{-[\ln(\mu_P/n)]^2/2s^2\}$ was used, in which $\langle\mu_P\rangle = n \cdot \exp(s^2)$ is the average magnetic moment per particle and $s^2$ is the $\ln(\mu_P)$ dispersion. The lognormal distribution function takes into account the distribution of values of the particle magnetic moments over the particle size. During the processing of the experimental *M(H)* dependences, the $N_P$ and $s$ values were fixed (not changed at different temperatures) and only the parameters $n$ (actually, $\langle\mu_P\rangle$) and $\chi_{PM}$ varied with temperature. The best agreement was achieved at $N_P = 2.48 \cdot 10^{18}$ particles per gram of the sample mass and $s = 0.27$. The results of the best fitting of the *M(H)* dependences are shown by solid lines in Fig. 4a. The partial components of the fitting curves are shown in Fig. 4b. The first term in Eq. (2) is designated as $M_{FM}(H)$ and the second term, as $M_{PM}(H)$.

The temperature evolution of the $\langle\mu_P\rangle$ value, which determines the behavior of the $M_{FM}(H)$ dependence, is illustrated in Fig. 5 (the $\langle\mu_P\rangle$ values are given in Bohr magnetons $\mu_B$). The experimental $\langle\mu_P\rangle(T)$ dependence is well described by the equation

$$\langle\mu_P\rangle = \langle\mu_P\rangle_{(T=0)} \cdot (1 - B \cdot T^\alpha) \qquad (3)$$

at $\alpha = 1.5 \pm 0.1$ ($B$ is the constant). For ferrimagnetic nanoparticles, different authors observed the $\alpha$ values both different from the classical Bloch value 3/2 [49,50] and equal to it [9,14,51]. According to the data presented in Fig. 5, we have $\langle\mu_P\rangle(T=0) = 296\ \mu_B$. This value corresponds to the saturation magnetization $M_{S\_FM}(T = 0)$ determined as $M_{S\_FM} = <\mu_P> \cdot N_P$, according to which $M_{S\_FM}(T = 0) \approx 6.8$ emu/g.

The SPM behavior is determined by the magnetic moment of the ferrimagnetically ordered particle core, while the magnetic response from atoms in the surface layer $d_{md}$ is naturally identified with the second term of Eq. (2). Figure 5 shows the temperature dependence of $\chi_{PM}$. It can be seen that this dependence changes with temperature as $1/T$, which allows us to consider the second term of Eq. (2) as a magnetic response from the PM subsystem. The established dependence

$$\chi_{PM}(T) = \chi_{PM(T=0)} \cdot (1/T) \qquad (4)$$

indicates that, at high temperatures, the PM subsystem is unrelated to the subsystem of ferrimagnetically ordered particle cores.

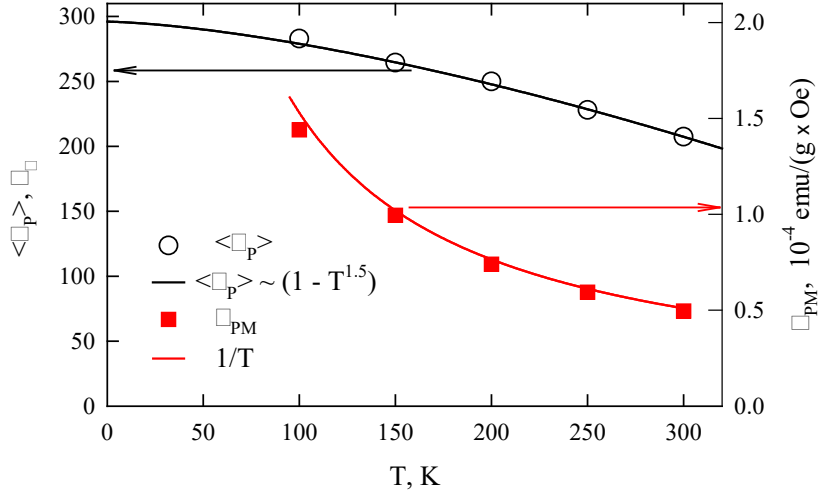

Fig. 5. Temperature evolution of average particle magnetic moment $<\mu_P>$ (the ordinate axis on the left) and $\chi_{PM}$ (the ordinate axis on the right). Solid curves are calculated using Eqs. (3) and (4).

Thus, the processing of the magnetization isotherms revealed the contributions of two magnetic subsystems, which can be attributed to (i) the ferrimagnetically ordered $NiFe_2O_4$ particle core (the SPM behavior) and (ii) the surface spins in the PM state at high temperatures. It is interesting to compare the temperature dependence of the magnetization with the calculated curve in the vicinity of the characteristic temperature $T_{max}$ in order to determine which maximum corresponds to a particular subsystem. Figure 6a shows the experimental $M(T)_{ZFC}$ dependence in a field of $H = 100$ Oe and the calculated $M_{FM}(T)$, $M_{PM}(T)$, and $M_{tot}(T)$ dependences. According to Eq. (4), we obviously have $M_{PM}(T) = \chi_{PM}(T) \cdot H = \chi_{PM(T=0)} \cdot H \cdot (1/T)$ at $H = 100$ Oe. The $M_{FM}(T)$ dependence corresponds to the first term of Eq. (2) with the above parameters $s$ and $N_P$ at $H = 100$ Oe; in this case, the $<\mu_P>(T)$ dependence is determined by Eq. (3). Then,

$$M_{tot}(T)_{H=100\,Oe} = M_{FM}(T)_{H=100\,Oe} + M_{PM}(T)_{H=100\,Oe}. \qquad (5)$$

Good agreement between the $M(T)_{ZFC}$ and $M_{tot}(T)$ dependences is observed over almost the entire temperature range; upon approaching the temperature $T_{max}$, the calculated $M_{tot}(T)$ dependence starts deviating from the experiment at ~10 K above the temperature $T_{max}$ (Fig. 6a). The comparison of the change in the magnetization of the high temperature bell and the $M_{FM}(T)$ contribution to the total magnetization $M_{tot}(T)$ shows that $T_{max}$ is the temperature of the SPM blocking of the magnetic moments of the particle cores. The surface spin subsystem freezes at a lower temperature and, in fact, the $M_{PM}(T)$ dependence or its part must change its behavior near $T_S$ from the growth to the drop with decreasing temperature.

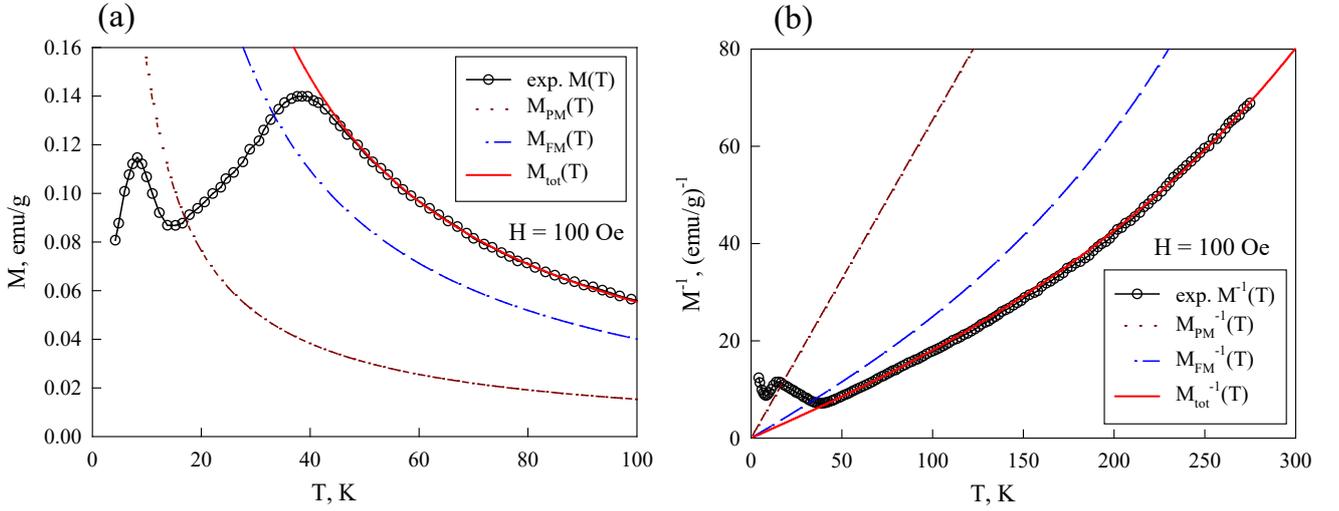

Fig. 6. (a) ZFC temperature dependence of the magnetization in a field of $H = 100$ Oe. The solid line is the $M_{tot}(T)$ dependence for the SPM state built using Eq. (5) on the basis of the processed $M(H)$ dependences. Dashed lines are the contributions $M_{PM}(T)$ of the particle magnetic moments and $M_{FM}(T)$ of the PM subsystem (the terms of Eq. (6)).

We note one more feature in the behavior of the $M(T)$ dependences at fairly high temperatures. The above-discussed experimental and calculated temperature dependences of the magnetization in the coordinates ($M^{-1}$, $T$) are presented in Fig. 6b. The experimental $M(T)$ dependence in these coordinates is not a linear function, as could be expected for the SPM state of the magnetic moments of particles and the PM state of the surface spins at temperatures higher than the SPM blocking temperature. The calculated $M(H = 1$ kOe, $T)$ dependence describes well the nonlinear behavior that can be seen in Fig. 6b. Of the two terms of Eq. (5), only the first term $M_{FM}(T)_{H = 100\ Oe}$ behaves nonlinearly in the coordinates ($M^{-1}$, $T$). This behavior of the magnetization, which deviates from the $1/T$ law, is caused by the temperature dependence of the average magnetic moment of particles (see Eq. (3) and Fig. 5).

*3.4 AC Susceptibility*

Figure 7 presents the temperature dependences of the real ($\chi'$) and imaginary ($\chi''$) parts of the magnetic susceptibility (the ac field amplitude is 2 Oe) and the dc susceptibility $\chi_{DC}(T) = M(T)_{ZFC}/H$ in a magnetic field of $H = 2$ Oe for the investigated sample. The presented curves exhibit some interesting features that characterize the dynamics of the nanoparticle magnetic moments. First, this is the presence of low- and high-temperature susceptibility maxima at $T = T_S$ and $T = T_{max}$, respectively, observed also in the $M(T)_{ZFC}$ dependences (Fig. 3). Second, the $T_S$ and $T_{max}$ positions shift towards higher temperatures with increasing ac field frequency, which is typical of the SPM blocking processes and spin-glass-like transitions. The imaginary part of the susceptibility, in addition to the two pronounced maxima, has a bright feature of the shape of the $\chi''(T)$ curve in the form of a shoulder between $T_S$ and $T_{max}$. The shape of this plateau-like feature also depends on frequency. It can be assumed that the temperature range of the $\chi''(T)$ shoulder corresponds to the extended temperature drop in the $\chi'(T)$ dependences.

As was noted in Subsection 3.3, the plateau-like behavior of the $M(T)_{ZFC}$ dependence in fields of 0.5–1.5 kOe near the temperature $T_{max}$ indirectly indicates the presence of fairly strong interparticle magnetic interactions. These interactions between the nanoparticle magnetic moments are most clearly manifested in the $\chi''(T)$ dependence as a characteristic shoulder to the left of the temperature $T_{max}$. Previously, we showed that this shoulder arises due to the formation of a collective state of the magnetic moments on the nanoparticle surface [26]. In other words, neighboring particles contact with the formation of a common magnetic subsystem in a nanoparticle cluster, the sizes of which depend on the degree of the magnetic interactions [26,52].

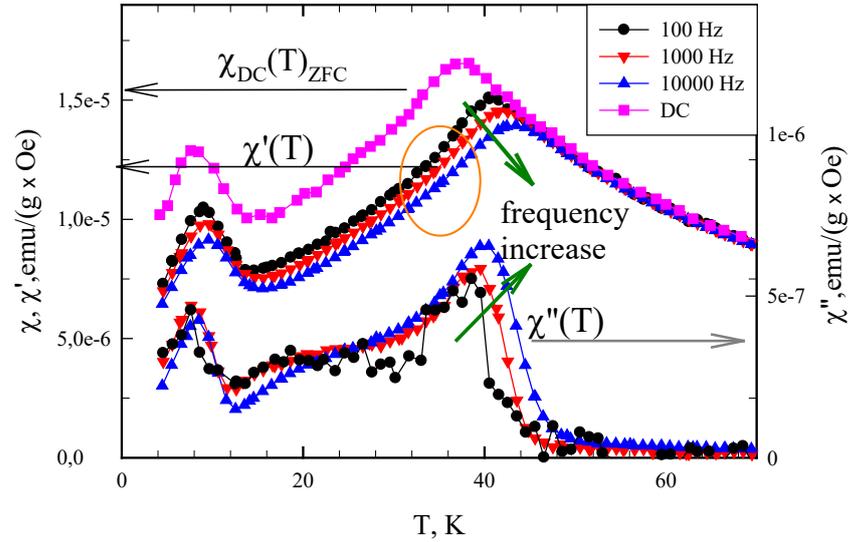

Fig. 7. Temperature dependences of the real ($\chi'$) and imaginary ($\chi''$) parts of the magnetic susceptibility (the ac field amplitude is 2 Oe) in the frequency range of 100–10000 Hz and dc susceptibility $\chi_{DC}(T) = M(T)_{ZFC}/H$ in a magnetic field of $H = 2$ Oe.

It is generally accepted that the imaginary part of the magnetic susceptibility characterizes the energy dissipation processes during blocking (or freezing) of the magnetic moments. Blocking of the magnetic moment along the nanoparticle anisotropy axis slows down the reversal of the uncompensated magnetic moment of a nanoparticle, which leads to absorption of the energy, i.e., the magnetic energy loss. If the freezing process components are of different nature, their characteristic temperatures will be different. Then, analysis of the $\chi''(T)$ dependence makes it possible to separate contributions of different natures to the blocking of the magnetic moments of ultrafine nanoparticles. The mathematical modeling of the process under study can be performed analytically using the equation

$$\chi'' = c \cdot \psi(f, \gamma) \cdot \sum_{i=1}^{n_K} q_i \frac{\theta^*}{\theta_{0,i}^2} \exp\left(-\frac{\ln^2\left(\theta^*/\theta_{0,i}^2\right)}{2\delta_\theta^2}\right) \quad (6)$$

This equation takes into account the existence of several magnetic subsystems ($n_K$) in the sample and the log-normal nanoparticle size distribution with the standard deviation $\delta_\theta$; $\theta_0 = \dfrac{E_A}{k_B}$. The use of the lognormal distribution function in fitting the temperature dependence of the magnetic susceptibility makes it possible to take into account the available nanoparticle size distribution and, consequently, the effect of this distribution on the magnetic properties of the sample. In this case, three characteristic bell-shaped features (two peaks and a shoulder) can be seen in the temperature dependence of the imaginary part of the susceptibility, each corresponding to a separate magnetic subsystem. Therefore, in Eq. (6), we have the right to take the number of magnetic subsystems to be $n_K = 3$; their weighting factor $q_i$ was varied during the modeling, as the distribution width $\delta_\theta$. All the rest parameters were described in detail in [26,53,54]. This approach, with independence of the weight parameters $q_i$ on frequency, suggests that the energy dissipation occurs in each subsystem independently; i.e., the blocking processes belong to different magnetic subsystems. The result of approximation of the $\chi''(T)$ dependence is shown in Fig. 8 and the numerical data are given in Table 2.

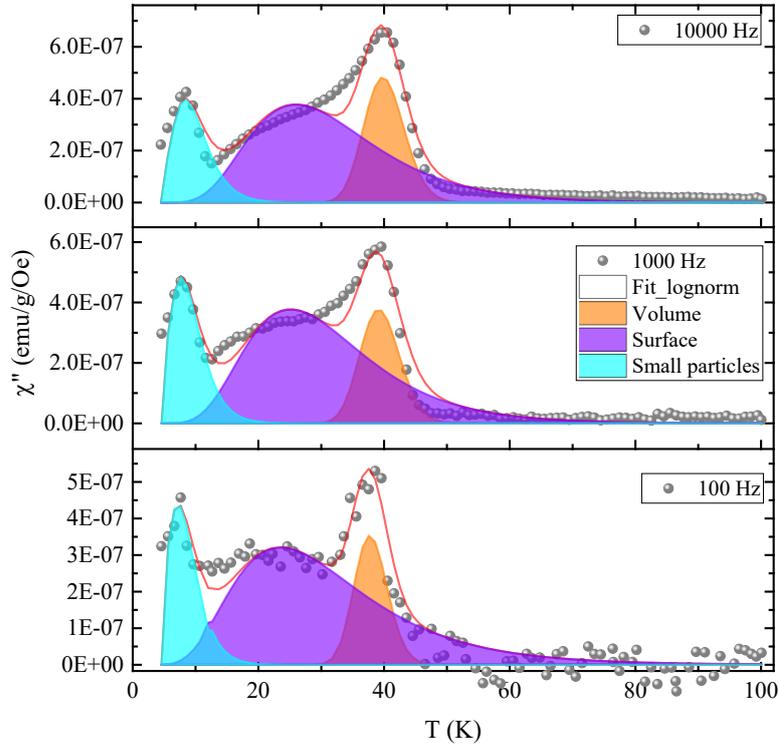

Fig. 8. Temperature dependences of the imaginary part of variable magnetic susceptibility $\chi''$ at frequencies of 100, 1000, and 10000 Hz for the nanoparticles under study. Dots show the experimental data and the solid line, the result of fitting. Partial components are shown as shaded areas.

Thus, instead of the preliminary hypothesis about two magnetic subsystems, which is based on the two pronounced maxima observed in the temperature dependences of the dc magnetization, the data on the imaginary part of the magnetic susceptibility speak already about three magnetic subsystems. Let us pay attention to the ratio between the contributions of the obtained components of the dissipation process, which is common for the system (Table 2). The smallest contribution can be attributed to the subsystem of individual magnetic moments of atoms (highlighted in cyan in Fig. 8) and the corresponding fitting parameters (the energy is the area under the bell) are designated as $E_{small}$ and "small". We believe that, in the vicinity of $T_S$, freezing of the atomic magnetic moments of disordered fine particles occurs. The most significant contribution (specifically, the absolute $\chi''$ value and the wide temperature range of the bell in the $\chi''(T)$ dependence) to the dissipation process is made by the subsystem associated above with the correlated surface spins in particles (see Fig. 8). It corresponds to the shoulder in the total $\chi''(T)$ dependence; the partial contribution of this subsystem is highlighted in purple in Fig. 8 and the parameters corresponding to it in Table 2 are designated as $E_{shell}$ and "shell". The subsystem of the particle magnetic moments passes to the blocked state (or freezes) in the vicinity of the high-temperature maximum; in Table 2, the parameters corresponding to this subsystem are designated as $E_{core}$ and "core".

The anisotropy energies estimated at the approximation of the $\chi''(T)$ dependence for each magnetic subsystem are given in Table 2. Here, we also note that, in the sample under study, the core/shell area ratio depends significantly on the frequency of an applied field, while the fraction of fine particles remains constant (the last column in Table 2). This is undoubtedly related to the manifestation of the interparticle interactions and decoupling of the magnetic subsystems of the bulk part of particles and the surface correlated spins. The frequency dependence of the core/shell volume ratio can be interpreted as a manifestation of the interaction between the subsystems of surface spins and particle magnetic moments [26]. This is probably due to the fact that the frequency of precession of the magnetic moments in the surface layer varies with distance to the particle center. Thus, changing the frequency of an external field, one can affect the interaction between these magnetic subsystems. Note that, as shown below, the aforesaid is valid for particles at least 4 nm in size.

Table 2. Results of fitting of the temperature dependences of the imaginary part of the magnetic susceptibility. $E_{core}$ is the magnetic energy of nanoparticles, $E_{shell}$ is the energy of freezing of the surface magnetic moments, and $E_{small}$ is the energy of freezing of the magnetic moments of fine particles.

| $f$, Hz | $E_{core}$ ($\times 10^{-15}$ erg) | $E_{shell}$ ($\times 10^{-15}$ erg) | $E_{small}$ ($\times 10^{-15}$ erg) | Area core | Area shell | Area small |
|---|---|---|---|---|---|---|
| 100 | 21±2 | 16±1 | 3.5±0.2 | 0.16 | 0.66 | 0.18 |
| 1000 | 21±2 | 16±1 | 3.5±0.2 | 0.19 | 0.62 | 0.19 |
| 10000 | 21±2 | 16±1 | 3.5±0.2 | 0.23 | 0.60 | 0.17 |

## 4. Discussion

The magnetization data obtained allow us to make some quantitative estimations of the magnetic contributions to the magnetic moment of NiFe$_2$O$_4$ nanoparticles. The $M_{S\_FM}$ value obtained at $T = 0$ in Subsection 3.3 is 6.8 emu/g, which is much less than the corresponding value for bulk NiFe$_2$O$_4$ ($M_{Sbulk} \approx 50$ emu/g [11]). This is indicative of a high degree of defectiveness of nanoparticles. Obviously, the $M_{S\_FM}/M_{Sbulk}$ ratio corresponds to the particle volume fraction with the ferrimagnetic ordering. In other words, if we divide a particle with volume $V$ into a ferrimagnetically ordered core with volume $V_{CORE}$ and a disordered surface layer with volume $V_{SHELL}$ ($V = V_{CORE} + V_{SHELL}$), we have $M_{S\_FM}/M_{Sbulk} = V_{CORE}/V$. The $M_{S\_FM}/M_{Sbulk}$ ratio is merely ~0.14. The thickness $d_{md}$ of the magnetically dead layer can be estimated as

$$M_{S\_FM} = M_{Sbulk} \cdot (1 - 2 \cdot d_{md}/d)^3. \quad (7)$$

Using Eq. (7), at $d = <d> \approx 4.0$ nm, we obtain $d_{md} \approx 1.0$ nm. Obviously, at this $d_{md}$ value, a particle smaller than 2 nm will be completely disordered and nonmagnetic; moreover, it makes sense to speak about a significant magnetic moment of a particle only if the ferrimagnetically ordered core has a volume of several NiFe$_2$O$_4$ unit cells, i.e., for particles with a size of $d \geq 4$ nm.

Let us establish a relation between the magnetic moment distribution function $f(\mu_P)$ obtained by processing of the $M(H)$ dependences and the particle size distribution $f(d)$. In the case of coarse (and perfectly ordered) particles, we have $\mu_P \sim V \sim d^3$; then, the $M_{Sbulk}$ value for NiFe$_2$O$_4$ corresponds to $2.3\mu_B$ per formula unit. The factor of proportionality between $\mu_P$ and $d$ can be obtained by calculating the number of formula units in a particle basing on average distance $d_{am}$ between magnetically active atoms. The distances between Ni–Ni, Ni–Fe, and Fe–Fe atoms in the NiFe$_2$O$_4$ structure are close enough, so a value of $d_{am} \approx 0.35$ nm can be used [55]. The number $N_{FU}$ of formula units in a cubic particle can be calculated as $N_{FU} \approx \frac{1}{3} \cdot [(d/d_{am}) + 1]^3$; consequently, the magnetic moment of a particle can be approximately estimated from the relation $\mu_P \approx N_{FU} \cdot 2.3 \mu_B$. Using these expressions, it can be found that, at $d \geq 2$ nm, the magnetic moment of one cubic nanometer of ordered NiFe$_2$O$_4$ is approximately $18\mu_B$. Taking into account the difference between the actual particle size $d$ and the size of the magnetically ordered core ($d - 2 \cdot d_{md}$), the relation between $\mu_P$ and $d$ can be written as

$$\mu_P \approx (\tfrac{1}{3}) \cdot [(d - 2 \cdot d_{md})/d_{am} + 1]^3 \cdot 2.3\ \mu_B. \quad (8)$$

Figure 9a shows, together with the particle size distribution histogram built using the TEM data, the distribution $f(d + 2 \cdot d_{md})$ at $d_{md} = 1.25$ nm (solid curve) obtained from the $f(\mu_P)$ function using Eq. (8). The $f(\mu_P)$ function at $T = 0$ (<$\mu_P$> = 298 $\mu_B$, see Subsection 3.3) is presented in the inset to Fig. 9a. It was built using a value of $s = 0.27$ obtained by processing of the $M(H)$ dependences, as well as the <$\mu_P$> value corresponding to <$\mu_P$>$_{(T = 0)}$ = 296 $\mu_B$[1] at $n = 285$ (see Subsection 3.3). We can state good agreement between the real size distribution (the TEM data) and the distribution $f(d + 2 \cdot d_{md})$, taking into account the strongest influence of the disordered surface layer in fine particles. For comparison, Fig. 9a shows the function $f(d + 2 \cdot d_{md})$ at $d_{md} = 0$ (dashed curve), which does not agree with the size distribution built from the TEM data, since it does not take into account the presence of a disordered magnetically dead layer. The unshaded columns in the histogram (Fig. 9a) correspond to the particles smaller than 4 nm, in which a ferrimagnetically ordered core has not formed; at the sizes corresponding to the shaded columns, the magnetically ordered core is about $d - 2 \cdot d_{md}$ in size. Since the size of the NiFe$_2$O$_4$ unit cell is about 8.35 Å [6] and contains eight formula units, there are about 10 unit cells in the volume of a sphere 2 nm in diameter. Thus, taking into account the presence of defects in particles of this size, the long-range magnetic ordering seems improbable. Naturally, such particles go to the source of a fairly strong paramagnetic contribution, which, according to the magnetic measurement data (Figs. 4b and 5), manifests itself at high temperatures.

---

[1] The value <$\mu_P$>$_{(T = 0)}$ = 296 $\mu_B$ corresponds to a volume of ~16 nm$^3$ or a magnetically ordered particle core size of ~2.6 nm. Since we have $d_{md} \sim 1$ nm, the $d$ value will be ~4.5 nm, which is similar to the average particle size (~4 nm) found from to the TEM data.

Analysis of the X-ray absorption spectroscopy (XAS) and X-ray magnetic circular dichroism (XMCD) data on the MFe$_2$O$_4$ spinel films [56] showed that the magnetically dead layers contain a significant number of vacancies in tetrahedral sites of the spinel structure, which may be a microscopic reason for the ferrimagnetic order violation in thin films. In other words, the magnetically dead layer is formed by structural defects in the packing of tetrahedral and octahedral sites of the spinel structure, which is also observed in our case by Mössbauer spectroscopy (see Table 1). In addition, we observe the cation redistribution between tetrahedral and octahedral sites, which, according to our estimates, should lead to an increase in the magnetic moment. However, the results of the magnetic measurements did not show this, which suggests a partial loss of magnetic order in defective areas (fine nanoparticles).

As noted in [57,58], the structural disorder promotes the formation of spin glass-like magnetic structures on the nanoparticle surface. Therefore, it is interesting to compare the dead magnetic layer thickness in nanoparticles estimated from the magnetization data with the results of the structural/morphological analysis. Here, the Mössbauer spectroscopy data can be used, since this technique is sensitive to the local environment of iron cations in the structure and to the possible occurrence of stronger distorted defect states on the particle surface. In [25], using this approach, we demonstrated that a dense core of ferrihydrite nanoparticles is only formed at a particle size of more than 2 nm. Finer particles represent some amorphous formations, which is reflected in the magnetic properties of this material.

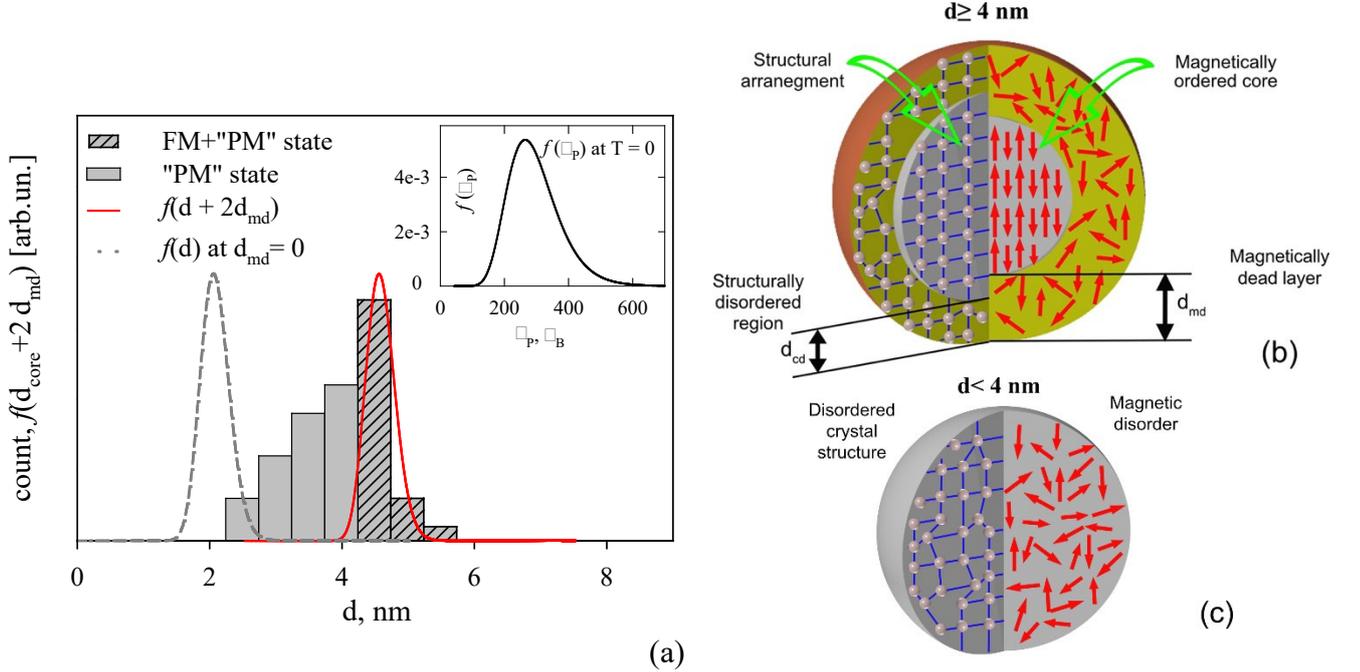

Fig. 9. (a) Comparison of the nanoparticle distribution histogram and the distribution of magnetically ordered nanoparticles over blocking temperatures. Schematic of the interplay between the morphology and magnetic ordering in NiFe$_2$O$_4$ nanoparticles at (b) $d \geq 4$ nm and (c) $d < 4$ nm. The left-hand side of the particle shows the distribution of structurally ordered and defective areas in it. The right-hand side shows magnetically ordered (core) and disordered (shell) areas.

For NiFe$_2$O$_4$ nanoparticles, we obtained (see Table 1) that, in the blocked state of the nanoparticle magnetic moments, at 4 K, 35% of iron atoms (in tetrahedral and octahedral sites in total) have the Mössbauer parameters close to those for bulk NiFe$_2$O$_4$ [59]. Therefore, the local environment of these atoms can be considered similar to that in spinel crystals. Then, we can easily estimate the diameter $d_{cryst}$ of the particle area in which a more ordered crystal structure is preserved. Since the number of atoms is proportional to the volume in which they are distributed, we obtain the equation $<d>^3 \cdot 0.35 = d_{cryst}^3$. Then, at $<d>$ = 4 nm, we have $d_{cryst} \approx 2.8$ nm; therefore, the crystallographically disordered layer will have a thickness of $d_{cd} \approx 0.6$ nm, which is somewhat smaller than the magnetically dead layer thickness determined by the magnetization analysis.

In this regard, it is interesting to compare the results obtained with the experimental ac susceptibility data for the sample under study. The processing of the $\chi''(T)$ dependences allowed us to draw some intriguing conclusions. First, by analyzing the magnetic energy dissipation processes occurring upon blocking of the magnetic moment, one can easily separate the processes of different nature, specifically, the energy loss as a result of (i) blocking of the total magnetic moment from the magnetically ordered particle core, (ii) blocking of the correlated spins, and (iii) ordering (spin glass) in the finest particles. This is brightly illustrated by the modeled $\chi''(T)$ dependences (see Subsection 3.4 and Fig. 8). Second, the modeling allows us to estimate the weight contribution to the energy loss due to each of these processes, which will be expressed by the area under the corresponding component. Quantitatively, we observe full agreement with the magnetic measurement and Mössbauer spectroscopy data. As noted above, the greatest contribution comes from the energy dissipation upon blocking of the correlated spins on the particle cluster surface, which is consistent with the results reported in [26]. This confirms that most of the magnetic moments of magnetoactive atoms are located in the magnetically dead layer at the nanoparticle surface.

Based on the results obtained, we can propose a model for the magnetic structure formation in such ultrafine ferrite nanoparticles, taking into account the structural disorder of atoms in them (see the schematic in Figs. 9b, 9c). According to our ideas, magnetic cations have a regular local environment in the nanoparticle core region, while the magnetic structure is preserved in the region with a smaller size. Hence, the disordered, but correlated magnetic moments on the surface enhance the effect of structural disorder and the magnetically dead layer penetrates into the particle region with a more regular crystal structure. Thus, we observe the mutual enhancement of these two effects in ultrafine nanoparticles. This behavior can be explained by the fact that the magnetic structure of a solid is determined by the crystal structure, with its symmetry and positioning of atoms in a lattice. Since the crystal structure in such fine nanoparticles includes a great number of defects (cation vacancies and broken bonds), the distribution of the magnetic exchange interactions will be strongly defective as well. The magnetic order is apparently more sensitive to defects, which is manifested in its faster disappearance when moving away from the particle center. In addition, it should be noted that the observed discrepancy is relatively small and amounts to only one atomic layer.

**5. Conclusions**

Thus, the defective crystal and magnetic structures of ultrafine $NiFe_2O_4$ nanoparticles were analyzed by comparing the TEM, $^{57}Fe$ Mössbauer effect, and dc magnetization and ac susceptibility measurement data. The estimation of the magnetic characteristics (saturation magnetization) at the known magnetic structure of the material showed that most of the magnetic moments of atoms are in the disordered state. According to the magnetic measurement data, the thickness of the magnetically dead layer in nanoparticles is $d_{md} \approx 1$ nm at a nanoparticle diameter equal to the average size $<d>$ = 4 nm found from the distribution function. This estimate is consistent with the data of $^{57}Fe$ Mössbauer spectroscopy, which is highly sensitive to minor changes in the local $^{57}Fe$ crystallographic and magnetic environment. Using the data obtained, the dead magnetic layer thickness in particles of similar size was found to be $d_{md} \approx 0.6$ nm. The resulting discrepancy was attributed to the fact that the magnetic order is maintained in ultrafine nanoparticles at a smaller distance to the nanoparticle center as compared with the crystallographic order. In our opinion, this can have two causes. First, the correlated magnetic moments on the particle surfaces enhance the effect of the structural disorder and the magnetically dead layer penetrates into the particle region with a more regular crystal structure. Second, the structural disorder of the particle surface violates the ferrimagnetic ordering of the particle core. The magnetic disorder spreads to greater depths. Therefore, the presence of a disordered surface layer in nanoparticles determines the core–shell type of the magnetic structure. The estimated thickness of the magnetically dead layer provides a simple explanation for the presence of an additional frequency-dependent maximum in the temperature dependences of the magnetization and susceptibility at $T_S \approx 8$ K. This maximum corresponds to the spin-glass state of the magnetic moments of atoms in the nanoparticles in the core of which the ferrimagnetic ordering cannot form, i.e., in the particles smaller than 4 nm ($\sim 2d_{md} \approx 2$ nm). Thus, the combination of the *dc* and *ac* magnetization measurements with the Mossbauer $^{57}$Fe spectroscopy technique allowed us to reveal three magnetic subsystems in the investigated powder system of fine $NiFe_2O_4$ nanoparticles. Two of these systems, magnetic moments of the particle cores and correlated surface magnetic moments of the particle shells, interact with each other. In the high-temperature range, the surface spins, as well as the spins of atoms of the finest particles, behave paramagnetically, while the magnetic moments of the ferrimagnetically ordered particle cores exhibit the superparamagnetic behavior.

**Acknowledgements**

We thank M.S. Molokeev for the XRD measurements. The TEM, XRD, and ac susceptibility studies were carried out on the equipment of the Center for Collective Use, Krasnoyarsk Scientific Center, Siberian Branch of the Russian Academy of Sciences

**Funding**

This study was carried out within the State assignment of the Ministry of Science and Higher Education of the Russian Federation for the Kirensky Institute of Physics, Siberian Branch of the Russian Academy of Sciences. The synthesis of $NiFe_2O_4$ nanoparticles was carried out within the State assignment of the Ministry of Science and Higher Education of the Russian Federation for the Krasnoyarsk Scientific Center, Siberian Branch of the Russian Academy of Sciences.

**Data availability.**

Data is available on doi: 10.17632/7zftc43fph.1.